\documentclass[10pt,aps,prd,superscriptaddress,nofootinbib,nobibnotes,longbibliography,floatfix,twocolumn]{revtex4-2}

\usepackage{bm}
\usepackage{mathtools,
amsmath,
amssymb,
amsfonts,
mathrsfs,
chngcntr,
multirow}

\let\cc\corresponds
\let\corresponds\relax
\usepackage{mathabx}
\let\corresponds\cc

\usepackage[utf8]{inputenc}
\usepackage[T1]{fontenc}

\usepackage{soul}

\usepackage[dvipsnames]{xcolor}
\usepackage[unicode]{hyperref}
\hypersetup{colorlinks=true, citecolor=MidnightBlue,
            linkcolor=MidnightBlue, urlcolor=MidnightBlue, linktocpage=true}
\usepackage[normalem]{ulem}

\usepackage{graphicx}

\newcommand{\PRLsection}[1]{\noindent{\bf\emph{#1.---}}}

\begin{document}
\title{Screening of dipolar emission in two-scale Gauss-Bonnet gravity}

\author{Farid Thaalba}
\affiliation{Nottingham Centre of Gravity \& School of Mathematical Sciences, University of Nottingham, University Park, Nottingham NG7 2RD, United Kingdom}
\affiliation{SISSA, via Bonomea 265, 34136 Trieste, Italy \& INFN (Sez. Trieste), via Valerio 2, 34127 Trieste, Italy.}

\author{Leonardo Gualtieri}
\affiliation{Dipartimento di Fisica,  Universit\`a di Pisa  \& Sezione INFN Pisa, L. Bruno Pontecorvo 3, 56127 Pisa, Italy}

\author{Thomas P. Sotiriou}
\affiliation{Nottingham Centre of Gravity \& School of Mathematical Sciences, University of Nottingham, University Park, Nottingham NG7 2RD, United Kingdom}
\affiliation{School of Physics and Astronomy, University of Nottingham, University Park, Nottingham NG7 2RD, United Kingdom}

\author{Enrico Trincherini}
\affiliation{Scuola Normale Superiore, Piazza dei Cavalieri 7, 56126, Pisa, Italy \& INFN Sezione di Pisa, Largo Pontecorvo 3, 56127 Pisa, Italy}

\begin{abstract}
We study black holes in shift-symmetric scalar Gauss–Bonnet gravity extended by a cubic Galileon interaction with a distinct energy scale. Introducing this hierarchy profoundly modifies the theory’s phenomenology. The cubic interaction allows for smaller black holes, and can generate a screening mechanism near the horizon, making large Gauss–Bonnet couplings consistent with gravitational-wave bounds. Observable quantities such as the scalar charge, the innermost stable circular orbit, and its frequency are most affected for small black holes. The resulting multi-scale effective field theory remains technically natural and offers new avenues to probe gravity in the strong-field regime.
\end{abstract}

\maketitle

\PRLsection{Introduction}
Black holes (BHs) in general relativity (GR) are remarkably simple objects. In vacuum, stationary BHs are uniquely described by the Kerr metric~\cite{Carter:1971zc,Chrusciel:2012jk,Kerr:1963ud}, and, except for the electromagnetic field, Standard Model fields do not endow them with additional charges due to the ``no-hair'' theorems~\cite{Chase:1970omy,Hawking:1972qk,Hartle:1971qq,Teitelboim:1972ps,Bekenstein:1971hc,Bekenstein:1972ky}. These results motivate the Kerr hypothesis: all astrophysical BHs---except during highly dynamical stages---should be described by the Kerr geometry~\cite{Berti:2015itd,Sotiriou:2015pka,Cardoso:2016ryw,Herdeiro:2022yle}. Gravitational-wave observations now allow direct tests of this hypothesis and, more broadly, of the gravitational interaction in the strong-field regime~\cite{Gair:2012nm,Barack:2018yly,Yunes:2024lzm}.

Scalar-tensor theories provide a useful framework for exploring possible departures from the Kerr hypothesis~\cite{Berti:2015itd}. In most such theories, no-hair results continue to hold~\cite{Bekenstein:1995un,Sotiriou:2011dz,Hui:2012qt,Sotiriou:2015pka,Herdeiro:2015waa}. A well-known exception arises when a scalar field, $\phi$, couples to the Gauss--Bonnet (GB) invariant, $\mathcal{G}=R^{\mu \nu \rho \sigma} R_{\mu \nu \rho \sigma}-4 R^{\mu \nu} R_{\mu \nu}+R^{2}$~\cite{Kanti:1995vq,Pani:2009wy,Yunes:2011we,Sotiriou:2013qea}. In particular, a linear coupling $\alpha_{\text{GB}} \phi \mathcal{G}$, where $\alpha_{\text{GB}}$ is a coupling constant,
modifies the scalar equation,
\begin{align}
    \Box\phi = -\alpha_{\text{GB}}\,\mathcal{G},
\end{align}
so that the curvature of the BH spacetime sources the scalar field~\cite{Sotiriou:2013qea,Sotiriou:2014pfa}. 
This theory is invariant under a constant shift of the scalar field ($\phi \to \phi + \text{constant}$), which guarantees the scalar is massless and protected from radiative corrections. This property makes the theory particularly relevant, as current and future experiments probing physics at the horizon scale (the kilometer scale) will be able to test the existence of massless or very light fields~\cite{Maselli:2017cmm,Barausse:2020rsu,LISA:2022kgy,LISA:2024hlh,Abac:2025saz}.

Black holes in shift-symmetric scalar Gauss-Bonnet (sGB) gravity have two key properties~\cite{Sotiriou:2014pfa}. First, the scalar charge $Q=\lim_{r\to\infty}(-r^2\partial_r\phi)$, where $r$ is the areal radius,
is fixed with respect to the mass and spin of the BH and is not an independent parameter (notice that $Q$ is {\it not} a Noether charge). Additionally, BHs exist only above a minimum mass set by $\alpha_{\text{GB}}$. This raises a natural question from an effective field theory (EFT) perspective: can additional shift-symmetric interactions modify these properties?

Ref.~\cite{Saravani:2019xwx} showed that, in any shift-symmetric scalar-tensor theory with second-order equations, the scalar charge can be expressed as
\begin{align}
    \label{eq:charge}
    4\pi Q=\alpha_{\text{GB}}\int_{\cal H} n^\mu J^{\text{GB}}_{\mu}~, 
\end{align}
where $n^\mu$ is the horizon generator and $J^{\text{GB}}_{\mu}$ is the conserved current associated with shift symmetry. Hence, the contribution from additional interactions comes from $J^{\text{GB}}_{\mu}$ on the horizon and is suppressed by $\alpha_{\text{GB}}$ times the coupling constant of the interaction. 
Ref.~\cite{Thaalba:2022bnt} explored this quantitatively, by considering the action:
\begin{align}
        S &= \int \mathrm{d}^4 x\sqrt{-g}\left[\frac{M_{\text{Pl}}^2}{2}R+X+\frac{\alpha M_{\text{Pl}}}{\Lambda_{\text{GB}}^2}\phi\mathcal{G} \right. \nonumber \\
    &\left.+\frac{\gamma}{\Lambda_{\text{GB}}^2} G_{\mu\nu}\nabla^{\mu}\phi\nabla^{\nu}\phi
    +\frac{\sigma}{\Lambda_{\text{GB}}^3} X\Box\phi + \frac{\kappa}{\Lambda_{\text{GB}}^4} X^2\right]~,\label{eq:action_0}
\end{align}
where $g$ is the determinant of the metric, and $X=-\nabla_\mu\phi\nabla^\mu\phi/2$. We use 
natural units $c = \hbar=1$,\,\footnote{We use different conventions from~\cite{Thaalba:2022bnt}, were geometric units $G=c=1$ were used.} in which 
the Planck mass is $M_{\text{Pl}} = (8 \pi G)^{-1/2}$. The coupling constants $\alpha,~\gamma,~\sigma$, and $\kappa$ are dimensionless of $\mathcal{O}(1)$. 

The analysis of Ref.~\cite{Thaalba:2022bnt}, where all interactions share the same energy scale $\Lambda_{\text{GB}}$, confirmed that additional shift-symmetric interactions produce only mild quantitative changes in sGB BHs: the regularity condition persists, the minimum mass remains, and the scalar charge is only slightly affected.

A richer phenomenology emerges if different interactions are governed by distinct energy scales. For example, in~\cite{Eichhorn:2023iab}, an sGB gravity model with two scales is proposed to study potential imprints on supermassive BHs (see, however,~\cite{Thaalba:2025ljh}). A different approach is taken in~\cite{Hui:2021cpm}, where two scales in the Lagrangian are chosen such that the corresponding interactions become strongly coupled at the same energy. 
Another example is the EFT studied in~\cite{Noller:2019chl}, which aims to account for the accelerated expansion of the universe while producing observable imprints in gravitational waves from black hole binaries. A notable result from~\cite{Noller:2019chl} is the realization that the \textit{screening} mechanism (see e.g.~\cite{Babichev:2013usa} for a review on the Vainshtein screening) can significantly impact the theory's BH phenomenology.

In this paper, we consider a shift-symmetric action, but departing from~\cite{Thaalba:2022bnt}, we allow different interactions to have distinct energy scales, similar to the framework of~\cite{Noller:2019chl}. However, our approach is broader: rather than linking BH physics to dark energy, we aim to comprehensively analyse how a hierarchy of energy scales shapes the phenomenological properties of astrophysical BHs.
We find that the cubic Galileon interaction dramatically enlarges the viable parameter space of scalar Gauss–Bonnet gravity, enabling the existence of significantly smaller BHs. These solutions remain consistent with the EFT and with current observational constraints. The necessary ingredient is the screening mechanism that suppresses the effective Gauss–Bonnet coupling, thereby allowing substantially larger values of~$\alpha_{\text{GB}}$. We support this picture with analytical estimates and full numerical solutions. We further examine the impact on observables such as the location of the innermost stable circular orbit (ISCO) and its orbital frequency, finding that the most pronounced deviations arise for the smallest BHs.

\PRLsection{Two-Scale Model}
We shall focus exclusively on the $\sigma X\Box\phi$ term by setting $\gamma = \kappa = 0$, and allowing the characteristic scale, $\Lambda$, of the cubic interaction to be different from $\Lambda_{\text{GB}}$. As demonstrated in~\cite{Thaalba:2022bnt}, the $\sigma X\Box\phi$ term---among those considered in the action~\eqref{eq:action_0}---has the most significant impact on the phenomenology of sGB BHs. This term has been studied in various scenarios, ranging from the Galileon modification of gravity~\cite{Nicolis:2008in} to the large extra-dimensions DGP model~\cite{Luty:2003vm}. 

To factor out one of the three dimensionful scales ($M_\text{Pl}$, $\Lambda$, $\Lambda_\text{GB}$), we define a dimensionless scalar field $\hat{\phi}$ and a new scale $\hat{\Lambda}$ as 
\begin{align}
    \hat{\phi} \equiv \frac{\phi}{M_{\text{Pl}}}~, \quad \hat{\Lambda} \equiv \left(\frac{\Lambda^3}{M_{\text{Pl}}}\right)^{1/2}~.
\end{align}
Therefore, after renaming $\hat{\phi}$ back to $\phi$, the action becomes 
\begin{align}
    \label{eq:action_2}
    S &= M_{\text{Pl}}^2\int \mathrm{d}^4 x\sqrt{-g}\left[\frac{R}{2}+X+\frac{\alpha}{\Lambda^2_{\text{GB}}}\phi\mathcal{G}+\frac{\sigma}{\hat{\Lambda}^2} X\Box\phi\right]~.
\end{align}
The strong-coupling scales associated with the EFT interactions in~\eqref{eq:action_2}, at least around small backgrounds, are $\Lambda$ and $(\Lambda^2_\text{GB}M_\text{Pl})^{1/3}$. We remark, however, that the EFT is expected to lose its validity at a scale that is parametrically smaller than the strong coupling scale, to avoid violations of causality~\cite{Serra:2022pzl}.
As discussed in the following section, all these estimates depend on the distance from the source in the presence of a large scalar background (the screening effect).

In the following, we shall focus on energy scales for which the EFT correction may leave an observable imprint in the gravitational waveforms from binary BH coalescence. Therefore, we shall assume the scale of the sGB interaction, $\Lambda_\text{GB}$, to be comparable to the scale of a gravitational wave source, i.e., the inverse Schwarzschild radius, $1/r_h$, of a reference BH (which is expected to be in the $1-100$ km scale, for ground-based detectors). The scale of the cubic interaction, $\hat\Lambda$, will instead be assumed to be significantly smaller. We also note that, as pointed out in~\cite{Noller:2019chl}, the scalar GB interaction cannot be smaller (i.e., $\Lambda_\text{GB}$ larger) than the size of the quantum corrections it receives from the cubic Galileon interaction. Summarizing, we have the following hierarchy of scales:
\begin{equation}
    \label{eq:theo_bounds}
\hat\Lambda\ll\Lambda_\text{GB}\ll (\hat\Lambda M_\text{Pl}^2)^{1/3}~,
\end{equation}
and we shall consider the GB scale in the range $0.01\lesssim\Lambda_\text{GB}r_h\lesssim 10$. 

\PRLsection{Black holes and scalar charge}
We will first examine how the additional cubic Galileon interaction affects the scalar charge of a BH.
Let us first consider a static, spherically symmetric BH in pure shift-symmetric sGB gravity, without additional cubic interactions ($\sigma=0$). In the so-called decoupling limit, in which one neglects the back-reaction of the scalar field on the metric, the spacetime is given by the Schwarzschild metric, with the scalar field satisfying a regularity condition on the horizon that fixes the scalar charge to be $Q=(\alpha r_h)/(4 \Lambda_\text{GB}^2)$~\cite{Sotiriou:2013qea}.
In other words, the scalar hair is {\it secondary}~\cite{Kanti:1995vq}. The regularity condition that fixes the charge persists when back-reaction is turned on and only has solutions when the {\it existence condition}
$\Lambda_\text{GB}^4r_h^4-192\alpha^2>0$~is satisfied~\cite{Sotiriou:2013qea, Sotiriou:2014pfa}.
Thus, for any given value of the coupling constant $\alpha/\Lambda_\text{GB}^2$, the horizon radius (i.e., the BH mass) must be larger than a minimum value. The qualitative picture is the same for stationary, rotating BHs~\cite{Kleihaus:2015aje,Delgado:2020rev}. Including additional shift-symmetric interactions modifies but does not remove the regularity condition~\cite{Noller:2019chl, Thaalba:2022bnt}. When the cubic Galileon interaction $\sigma X\Box\phi$ is present, the existence condition becomes~\cite{Noller:2019chl, Thaalba:2022bnt}:
\begin{align}
  \label{eq:ext_cond}
   \hat{\Lambda }^4 \left(\Lambda _{\text{GB}}^4 r_h^6-192 \alpha ^2 r_h^2\right)-24 \alpha  \hat{\Lambda }^2 \sigma  \Lambda _{\text{GB}}^2 r_h^2 > 0~.
\end{align}
To study how the additional cubic term affects the scalar charge of the BH, we compute it in terms of the shift-symmetry conserved current $J_\mu^\text{GB}$, using Eq.~\eqref{eq:charge}. For ease of notation, we define the dimensionful coupling constant
$\alpha_{\text{GB}}=\alpha/\Lambda_\text{GB}^2$. 
As noted in~\cite{Creminelli:2020lxn}, the shift-symmetric current does not transform as a vector under diffeomorphisms, and it is not unique, even though its divergence $\nabla_\mu J^{\mu}_\text{GB} = \mathcal{G}$ is a genuine scalar. If the geometry possesses isometries, the current can be written as~\cite{Yale:2011usf, Creminelli:2020lxn}
\begin{align}
    J^{\mu}_\text{GB} = -2 P^{W\mu\nu}{}_{\rho}\Gamma^{\rho}{}_{\nu W}~, \quad P^{\mu\nu\rho\sigma} = \frac{\partial \mathcal{G}}{\partial R_{\mu\nu\rho\sigma}}~, \label{eq:Jiso}
\end{align}
where the index $W$ labels the direction of an isometry and is not summed over.

We write the metric of a  static, spherically symmetric BH in ingoing null coordinates $(v,r,\theta,\varphi)$, i.e.,
\begin{align}
    \mathrm{d}s^2 = f(r)\mathrm{d}v^2 + 2 h(r)\mathrm{d}r\mathrm{d}v + r^2\mathrm{d}\Omega^2~.
\end{align}
Eq.~\eqref{eq:Jiso}, with  $W=v$, gives
$J_{\mu}^{\text{GB}} = \left(J^{\text{GB}}_v,0,0,0\right)$ with
\begin{align}
    J^{\text{GB}}_v=-\frac{4 \left(f + h^2\right)f'}{r^2 h^3}~.
\end{align}
Finally, substituting in Eq.~\eqref{eq:charge}, we obtain a general expression for the scalar charge:
\begin{align}
    Q = -{\alpha_{\text{GB}}} \frac{4 \left(f + h^2\right)f'}{h^3}\bigg\vert_{r=r_h}~.
\label{eq:solQh}
\end{align}
In the decoupling limit, the spacetime is given by the Schwarzschild metric, thus $f(r)=-(1-2GM/r)$, $h(r)=1$, and
\begin{align}
    Q =  \frac{2\alpha_{\text{GB}}}{GM}~,
\label{eq:Qdec}
\end{align}
which agrees with~\cite{Sotiriou:2013qea}. 
Furthermore, we verified that the charge computed from Eq.~\eqref{eq:solQh}, taking into account back-reaction, agrees with the results of the numerical integration of the field equations of the theory with cubic interaction, Eq.~\eqref{eq:action_2}, which are given in Ref.~\cite{Thaalba:2022bnt}.

Let us now consider a near-horizon expansion of the spacetime metric. We carry out this calculation in Schwarzschild coordinates $(t,r,\theta,\varphi)$, 
\begin{align} 
    \mathrm{d}s^2 = -A(r)\mathrm{d}t^2 + \frac{1}{B(r)}\mathrm{d}r^2+ r^2\mathrm{d}\Omega^2~,\label{eq:AB}
\end{align}
which are related to the ingoing null coordinates by 
$\mathrm{d}v = \mathrm{d}t + \sqrt{A/B} ~\mathrm{d}r$, with $f=-A$, $g=\sqrt{A/B}$. In these coordinates, $g_{tt}$ and $g_{rr}^{-1}$ vanish on the horizon, leading to the following near-horizon expansion:
\begin{align}
    A(r) &= A_1(r-r_h)+\mathcal{O}(r-r_h)^2~, \nonumber \\
    B(r) &= B_1(r-r_h)+\mathcal{O}(r-r_h)^2~.
\end{align}
The $A_1$ coefficient is fixed by the requirement that the metric is asymptotically flat; we do not compute it explicitly at this stage. 
Since we choose the energy scale of the cubic interaction to be much smaller than the inverse of the characteristic BH length-scale, $\hat\Lambda\ll1/r_h$ (see Eq.~\eqref{eq:theo_bounds}), we shall expand Eq.~\eqref{eq:solQh} in the dimensionless parameter $\epsilon \equiv \hat{\Lambda} r_h$. We find that for $\alpha \sigma < 0$,\,\footnote{We do not consider $\alpha\sigma>0$ since the existence condition~\eqref{eq:ext_cond} is violated up to order $\mathcal{O}\left(\epsilon^2\right)$ in that case.}
\begin{align}
    \label{eq:Q_sigma}
    Q\rvert_{r=r_h}=\frac{4 \alpha  \sqrt{\frac{A_1}{r_h}}}{\Lambda _{\text{GB}}^2}\left(1+\frac{4\sqrt{6}\alpha^2\epsilon}{\sqrt{-\alpha\sigma}\Lambda_{\text{GB}}^3r_h^3}\right) + \mathcal{O}\left(\epsilon^2\right)~.
\end{align}
Therefore (assuming that $A_1$ remains finite as $\epsilon \to 0$), the cubic Galileon term appears to give a negligible contribution to the charge. 
\begin{figure}
    \centering
    \includegraphics[width=1\linewidth]{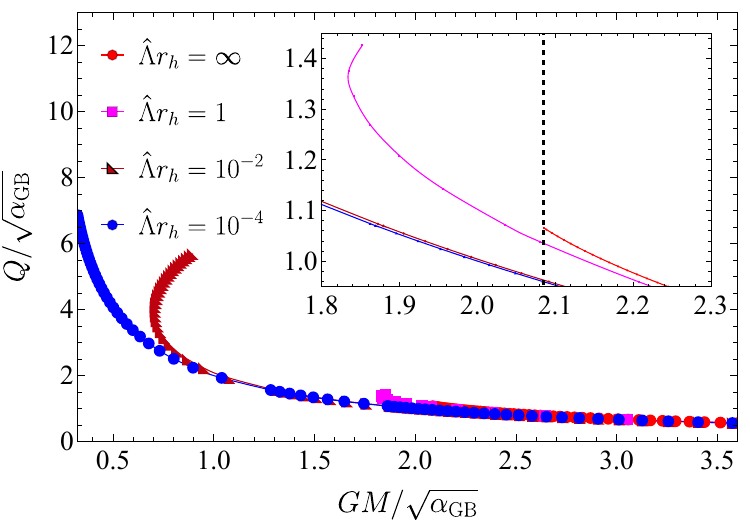} 
    \hspace{5mm}
    \caption{Normalized charge $Q/\sqrt{\alpha_{\text{GB}}}$ as a function of the normalized ADM mass $G M/\sqrt{\alpha_{\text{GB}}}$, for different values of the energy scale $\hat{\Lambda}$. The inset is a zoom-in of the region around the minimum mass for $\sigma=0$ indicated by the black dashed line.  We choose $\Lambda_{\text{GB}}=1/r_h$ with $\sigma=-1$.
    The presence of the cubic interaction does not significantly affect the values of the charge, but it does affect the allowed range of masses.}
    \label{fig:MQ}
\end{figure}
Considering the limitations of this perturbative analysis, we also confirm its findings numerically, by solving the full system of equations with back-reaction as discussed in~\cite{Sotiriou:2014pfa, Thaalba:2022bnt}. 
We present our numerical results in Fig.~\ref{fig:MQ}, where we show the scalar charge as a function of the ADM mass $M$, both normalized by $\alpha_\text{GB}$. We have chosen $\Lambda_\text{GB} = 1/r_h$ for illustrative purposes (observational bounds on $\Lambda_\text{GB}$ will be discussed below).

It is clear that, at least for masses larger than the minimum mass of pure sGB gravity ($GM/\sqrt{\alpha_\text{GB}}\simeq 2.1$, corresponding to the dashed line in the inset), the cubic Galileon interaction has little effect on the scalar charge. This is consistent with the perturbative analysis. On the other hand, the cubic interaction has a major effect on the minimum mass of BHs for a given $\alpha_{\text{GB}}$. Indeed, the presence of the additional interaction relaxes the bound on $\alpha_{\text{GB}}$ coming from the existence of BHs of a given mass. For $\alpha > 0$, $\sigma < 0$, and $\epsilon=\hat{\Lambda}r_h \ll \Lambda_{\text{GB}}r_h \sim 1$ this can be seen perturbatively as well: expanding Eq.~\eqref{eq:ext_cond} in $\epsilon \ll 1$ yields
\begin{align}
    0 < \alpha_\text{GB} < -\sigma\frac{r_h^2}{8 \epsilon^2} + \mathcal{O}\left(\epsilon^2\right)~,
\end{align}
hence smaller $\hat{\Lambda}$ leads to a larger upper bound on $\alpha_{\text{GB}}$. Note that this bound can be significantly larger than the bound of pure sGB gravity, $\alpha<r_h^2/\sqrt{192}$. 

\begin{figure}
    \centering
    \includegraphics[width=1\linewidth]{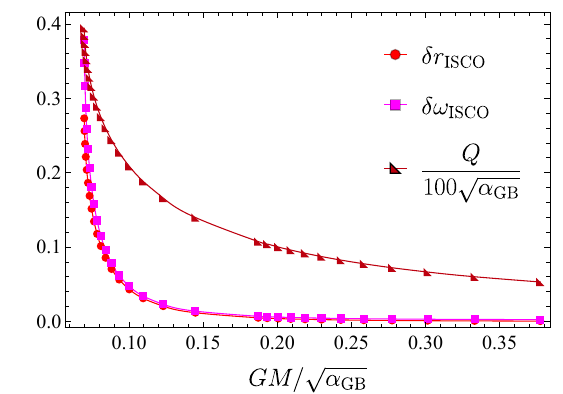}
    \caption{Relative difference in the location of the ISCO and the frequency at the ISCO between GR and the theory~\eqref{eq:action_2}, with $\Lambda_{\text{GB}}r_h = 0.1$ and $\hat{\Lambda}r_h=10^{-7}$. The deviations from GR are negligible, unless $Q/\sqrt{\alpha_\text{GB}}\gg 1$.}\label{fig:isco}
\end{figure}

To probe the effect of $\hat{\Lambda}$ on the BH spacetime, we consider the relative deviations for two phenomenologically relevant quantities: the location of the ISCO, $r_{\text{ISCO}}$, and the frequency at the ISCO times the gravitation radius, $\omega_{\text{ISCO}} r_g$, where $r_g = 2 G M$. In Fig.~\ref{fig:isco} we show the relative deviations from GR, defined as $\delta x = \left(x^{\text{GB}}-x^{\text{GR}}\right)/x^{\text{GR}}$, and the charge $Q$ normalized with the coupling constant. We have chosen $\Lambda_{\text{GB}}r_h = 0.1,~\hat{\Lambda}=10^{-7}$. Although the charge can take large values, the deviations only become significant for small BHs (which would not have existed without the cubic Galileon interaction).

\PRLsection{The screening effect}
We now show that the cubic Galileon interaction screens the sGB coupling near BHs. Working within the hierarchy of scales in Eq.~\eqref{eq:theo_bounds}, the background scalar $\phi_0(r)$ is sourced by the GB term, while perturbations $\varphi(t,r,\theta,\phi)$ about this background experience a renormalized kinetic term. This affects the dipolar emission in BH binary systems~\footnote{See~\cite{Dar:2018dra} for a numerical study on Vainshtein screening and emitted radiation from binary systems.}. 

Writing $\phi=\phi_0+\varphi$ and expanding the cubic interaction in Eq.~\eqref{eq:action_2} gives
\begin{align}
    \hat\sigma(\partial\phi)^2 \Box\phi
    =\big(\hat\sigma\,\Box\phi_0\big)(\partial\varphi)^2+\dots,\quad
    \hat\sigma\equiv\frac{\sigma}{\hat\Lambda^{2}}~,
\end{align}
so that perturbations have an effective kinetic term
\begin{align}
    -\frac{1}{2}\big(1+\hat\sigma\Box\phi_0\big)(\partial\varphi)^2\equiv
    -\frac{1}{2}\big(1+Z(r)\big)(\partial\varphi)^2~,
\end{align}
with $Z(r)\equiv \hat\sigma\Box\phi_0$. In regions where $Z\gg 1$ (the screened regime), the canonically normalized fluctuation $\varphi_c=\sqrt{Z}\,\varphi$ acquires a standard kinetic term, while the GB coupling to perturbations is suppressed~\footnote{For the same reason, all strong coupling scales are multiplied by the appropriate positive power of $Z$.}:
\begin{align}
    \mathcal{L}\supset \frac{\alpha}{\Lambda_{\text{GB}}^{2}}\,\varphi\,\mathcal{G}
    =\frac{\alpha}{\Lambda_{\text{GB}}^{2}\,Z^{1/2}}\,\varphi_c\,\mathcal{G}~.
\label{eq:newGBterm}
\end{align}
Equivalently, the effective GB scale for perturbations is
\begin{align}
    \Lambda_{\text{GB,eff}}^{2}=\Lambda_{\text{GB}}^{2}\,Z^{1/2}~.
\end{align}
Note that we are primarily interested in perturbations near the horizon of the BH. To estimate $Z$, we consider a static, spherically symmetric background and assume $Z\gg 1$ (to be verified \textit{a posteriori}). In this regime, the scalar equation schematically reads
\begin{align}
    \frac{\alpha}{\Lambda_{\text{GB}}^{2}}\mathcal{G}
    =\Box\phi_0+\frac{(\Box\phi_0)^2}{\hat\Lambda^{2}}
    \;\simeq\;\frac{(\Box\phi_0)^2}{\hat\Lambda^{2}}~,
\label{eq:eom_comp}
\end{align}
i.e., the cubic term dominates over the canonical one. For a Schwarzschild background, $\mathcal{G}\sim r_h^{2}/r^{6}$, hence $\mathcal{G}(r_h)\sim 1/r_h^{4}$. Using the definition of $Z$ and $\sigma=\mathcal{O}(1)$, we obtain
\begin{align}
Z^{2}(r)\sim\frac{\mathcal{G}}{\Lambda_{\text{GB}}^{2}\hat\Lambda^{2}}
\quad\Rightarrow\quad
Z(r_h)\sim\frac{1}{\hat\Lambda\,\Lambda_{\text{GB}}\,r_h^{2}}~.
\label{eq:Z_eom}
\end{align}
Current observations constrain the GB coupling for perturbations. In the absence of screening, the lower bound, coming from the observation of GW230529, implies $\Lambda_{\text{GB}}\gtrsim 10/r_h$~\cite{Sanger:2024axs}.
With screening, this translates to a bound on the \textit{effective} scale $\Lambda_{\text{GB,eff}} =\Lambda_{\text{GB}}Z^{1/4} \gtrsim 10/r_h$, which, together with Eq.~\eqref{eq:Z_eom}, yields the relation
\begin{align}
\left(\frac{\hat{\Lambda}}{\Lambda_{\text{GB}}^{3}r_h^{2}}\right)^{1/2}\lesssim 10^{-2}~.
\label{eq:hatLammbda}
\end{align}
This condition is compatible with Eq.~\eqref{eq:theo_bounds} and quantifies the range in which screening permits $\Lambda_{\text{GB}}$ values much lower than would be allowed without the cubic term. Table~\ref{tab:energyScales} lists upper bounds on $\hat{\Lambda}$ consistent with~\eqref{eq:hatLammbda} for representative values of the GB scale $\Lambda_{\text{GB}}$. 
\begin{table}[ht]
    \centering
    \begin{tabular}{cc|c|c}
       \multicolumn{2}{c|}{Energy Scales}& \multicolumn{1}{c|}{Normalizing Factor}\\
        $\Lambda_{\text{GB}} r_h$ & $\hat{\Lambda} r_h$ & $Z$\\
       \hline
          $10$   & $0.1$      &  $1$   \\
          $1$    & $10^{-4}$  &  $10^4$ \\
          $0.1$  & $10^{-7}$  &  $10^8$ \\
          $0.01$ & $10^{-10}$ &  $10^{12}$
    \end{tabular}
    \caption{Different energy scales consistent with the upper bound~\eqref{eq:hatLammbda}.}
    \label{tab:energyScales}
\end{table}

The main consequence of this mechanism is that of allowing smaller values of $\Lambda_{\text{GB}}$ to be consistent with observational bounds (e.g.\,\cite{Sanger:2024axs}), as in the presence of screening observations provide bounds on $\Lambda_{\text{GB,eff}}$. Since $Q$ $\propto$ $1/\Lambda_{\text{GB}}^{2}$ for fixed mass, this increases the possible values of the scalar charge, relative to pure sGB.

Another potentially observable consequence of screening is the following.  The interaction $X \Box \phi$ induces a kinetic mixing between the canonically normalized metric perturbation $h_c\equiv h/M_{\text{Pl}}$ and the scalar perturbation $\varphi_c$. A dimensional estimate gives~\cite{Noller:2019chl}
\begin{equation}
\label{eq:mixing_1}
M_{\text{Pl}}^{2}\,\partial h_c\,\frac{\partial\varphi_c}{Z^{1/2}}\,
\frac{(\partial\phi_0)^{2}}{\hat{\Lambda}^{2}}~.
\end{equation}
Since the background is a function solely of the radial coordinate $r$, taking a partial derivative has the effect, as far as orders of magnitude are considered, of dividing by $r$. We can therefore rewrite $\partial \phi_0$ as $r \partial^2 \phi_0$. Thus, Eq.~\eqref{eq:mixing_1} becomes:
\begin{align}
    M_{\text{Pl}}^2 \partial h_c \frac{\partial\varphi_c}{Z^{1/2}} \frac{\left(\partial^2 \phi_0\right)^2}{\hat{\Lambda}^4}r^2 \hat{\Lambda}^2 \sim  M_{\text{Pl}}^2 \partial h_c \partial\varphi_c \left(\frac{\hat{\Lambda}}{\Lambda_{\text{GB}}^3 r_h^2}\right)^{1/2}~.
\end{align}
To obtain the right-hand side, we substituted the definition of $Z$ and evaluated the expression near the horizon. Employing Eq.~\eqref{eq:hatLammbda}, the mixing is bounded by
\begin{align}
10^{-2}\,M_{\text{Pl}}^{2}\,\partial h_c\,\partial\varphi_c~,
\end{align}
i.e., at most an $\mathcal{O}(1\%)$ correction to the canonical kinetic terms. Notably, this bound is controlled by the observational requirement~\eqref{eq:hatLammbda} and is insensitive to individual choices of $\hat\Lambda$ and $\Lambda_{\text{GB}}$ within the allowed region.

\PRLsection{Discussion}
In this analysis, we focused on the effect of the additional cubic interaction on the scalar charge. We found that---for values of $GM/\sqrt{\alpha_\text{GB}}$ large enough that BHs exist in pure sGB gravity---the presence of the cubic interaction has a negligible effect on the scalar charge. However, even a small cubic term has a remarkable effect on the allowed range of BH masses, significantly reducing the minimum mass for a given value of the coupling $\alpha_{\text{GB}}$. Conversely, $\alpha_{\text{GB}}$ can be larger for a given minimum mass consistent with observations.

We also find that the cubic interaction induces a screening mechanism that affects all quantities related to scalar field perturbations, including the dipolar emission from BH binaries. As a consequence, both theoretical and observational bounds on $\Lambda_\text{GB}$ may be weaker than expected.
The screening mechanism occurs when scalar perturbations acquire a large non-canonical kinetic term that depends on the background field. This modifies the canonically normalized scalar field perturbation, leading to 
an effective GB coupling of the form $\Lambda_\text{GB}^2 Z^{1/2}$. 

Finally, we have estimated the kinetic mixing between scalar and metric perturbations, finding that it can be at most of order $1\%$, leading to similar deviations in the QNMs spectrum. If future detectors can measure such an effect on the QNMs, then these interactions can be further constrained. 

Although we have focused on the cubic Galileon interaction for concreteness, our results demonstrate a broader point: that additional interactions suppressed by an energy scale much lower than that of the Gauss-Bonnet coupling can lead to drastically different phenomenology, while satisfying the technical naturalness bound, avoiding strong coupling, and preserving consistency with current observational bounds. It would be interesting to consider other interactions, extend our analysis to rotating BHs, and quantify the effects these interactions could have on QNMs.

\PRLsection{Acknowledgments}
FT is supported by the INFN Post-doctoral research agreement No. 27076. LG acknowledges financial support from the EU Horizon 2020 Research and Innovation Programme under the Marie Sklodowska-Curie Grant Agreement No. 101007855. TPS acknowledges partial support from the STFC Consolidated Grant nos. ST/V005596/1 and ST/X000672/1. ET is partially supported by the Italian MIUR under contract 20223ANFHR (PRIN2022). 
\bibliography{biblio.bib}

\begin{thebibliography}{46}%
\makeatletter
\providecommand \@ifxundefined [1]{%
 \@ifx{#1\undefined}
}%
\providecommand \@ifnum [1]{%
 \ifnum #1\expandafter \@firstoftwo
 \else \expandafter \@secondoftwo
 \fi
}%
\providecommand \@ifx [1]{%
 \ifx #1\expandafter \@firstoftwo
 \else \expandafter \@secondoftwo
 \fi
}%
\providecommand \natexlab [1]{#1}%
\providecommand \enquote  [1]{``#1''}%
\providecommand \bibnamefont  [1]{#1}%
\providecommand \bibfnamefont [1]{#1}%
\providecommand \citenamefont [1]{#1}%
\providecommand \href@noop [0]{\@secondoftwo}%
\providecommand \href [0]{\begingroup \@sanitize@url \@href}%
\providecommand \@href[1]{\@@startlink{#1}\@@href}%
\providecommand \@@href[1]{\endgroup#1\@@endlink}%
\providecommand \@sanitize@url [0]{\catcode `\\12\catcode `\$12\catcode `\&12\catcode `\#12\catcode `\^12\catcode `\_12\catcode `\%12\relax}%
\providecommand \@@startlink[1]{}%
\providecommand \@@endlink[0]{}%
\providecommand \url  [0]{\begingroup\@sanitize@url \@url }%
\providecommand \@url [1]{\endgroup\@href {#1}{\urlprefix }}%
\providecommand \urlprefix  [0]{URL }%
\providecommand \Eprint [0]{\href }%
\providecommand \doibase [0]{https://doi.org/}%
\providecommand \selectlanguage [0]{\@gobble}%
\providecommand \bibinfo  [0]{\@secondoftwo}%
\providecommand \bibfield  [0]{\@secondoftwo}%
\providecommand \translation [1]{[#1]}%
\providecommand \BibitemOpen [0]{}%
\providecommand \bibitemStop [0]{}%
\providecommand \bibitemNoStop [0]{.\EOS\space}%
\providecommand \EOS [0]{\spacefactor3000\relax}%
\providecommand \BibitemShut  [1]{\csname bibitem#1\endcsname}%
\let\auto@bib@innerbib\@empty
\bibitem [{\citenamefont {Carter}(1971)}]{Carter:1971zc}%
  \BibitemOpen
  \bibfield  {author} {\bibinfo {author} {\bibfnamefont {B.}~\bibnamefont {Carter}},\ }\bibfield  {title} {\bibinfo {title} {{Axisymmetric Black Hole Has Only Two Degrees of Freedom}},\ }\href {https://doi.org/10.1103/PhysRevLett.26.331} {\bibfield  {journal} {\bibinfo  {journal} {Phys. Rev. Lett.}\ }\textbf {\bibinfo {volume} {26}},\ \bibinfo {pages} {331} (\bibinfo {year} {1971})}\BibitemShut {NoStop}%
\bibitem [{\citenamefont {Chrusciel}\ \emph {et~al.}(2012)\citenamefont {Chrusciel}, \citenamefont {Lopes~Costa},\ and\ \citenamefont {Heusler}}]{Chrusciel:2012jk}%
  \BibitemOpen
  \bibfield  {author} {\bibinfo {author} {\bibfnamefont {P.~T.}\ \bibnamefont {Chrusciel}}, \bibinfo {author} {\bibfnamefont {J.}~\bibnamefont {Lopes~Costa}},\ and\ \bibinfo {author} {\bibfnamefont {M.}~\bibnamefont {Heusler}},\ }\bibfield  {title} {\bibinfo {title} {{Stationary Black Holes: Uniqueness and Beyond}},\ }\href {https://doi.org/10.12942/lrr-2012-7} {\bibfield  {journal} {\bibinfo  {journal} {Living Rev. Rel.}\ }\textbf {\bibinfo {volume} {15}},\ \bibinfo {pages} {7} (\bibinfo {year} {2012})},\ \Eprint {https://arxiv.org/abs/1205.6112} {arXiv:1205.6112 [gr-qc]} \BibitemShut {NoStop}%
\bibitem [{\citenamefont {Kerr}(1963)}]{Kerr:1963ud}%
  \BibitemOpen
  \bibfield  {author} {\bibinfo {author} {\bibfnamefont {R.~P.}\ \bibnamefont {Kerr}},\ }\bibfield  {title} {\bibinfo {title} {{Gravitational field of a spinning mass as an example of algebraically special metrics}},\ }\href {https://doi.org/10.1103/PhysRevLett.11.237} {\bibfield  {journal} {\bibinfo  {journal} {Phys. Rev. Lett.}\ }\textbf {\bibinfo {volume} {11}},\ \bibinfo {pages} {237} (\bibinfo {year} {1963})}\BibitemShut {NoStop}%
\bibitem [{\citenamefont {Chase}(1970)}]{Chase:1970omy}%
  \BibitemOpen
  \bibfield  {author} {\bibinfo {author} {\bibfnamefont {J.~E.}\ \bibnamefont {Chase}},\ }\bibfield  {title} {\bibinfo {title} {{Event horizons in static scalar-vacuum space-times}},\ }\href {https://doi.org/10.1007/BF01646635} {\bibfield  {journal} {\bibinfo  {journal} {Commun. Math. Phys.}\ }\textbf {\bibinfo {volume} {19}},\ \bibinfo {pages} {276} (\bibinfo {year} {1970})}\BibitemShut {NoStop}%
\bibitem [{\citenamefont {Hawking}(1972)}]{Hawking:1972qk}%
  \BibitemOpen
  \bibfield  {author} {\bibinfo {author} {\bibfnamefont {S.~W.}\ \bibnamefont {Hawking}},\ }\bibfield  {title} {\bibinfo {title} {{Black holes in the Brans-Dicke theory of gravitation}},\ }\href {https://doi.org/10.1007/BF01877518} {\bibfield  {journal} {\bibinfo  {journal} {Commun. Math. Phys.}\ }\textbf {\bibinfo {volume} {25}},\ \bibinfo {pages} {167} (\bibinfo {year} {1972})}\BibitemShut {NoStop}%
\bibitem [{\citenamefont {Hartle}(1971)}]{Hartle:1971qq}%
  \BibitemOpen
  \bibfield  {author} {\bibinfo {author} {\bibfnamefont {J.~B.}\ \bibnamefont {Hartle}},\ }\bibfield  {title} {\bibinfo {title} {{Long-range neutrino forces exerted by kerr black holes}},\ }\href {https://doi.org/10.1103/PhysRevD.3.2938} {\bibfield  {journal} {\bibinfo  {journal} {Phys. Rev. D}\ }\textbf {\bibinfo {volume} {3}},\ \bibinfo {pages} {2938} (\bibinfo {year} {1971})}\BibitemShut {NoStop}%
\bibitem [{\citenamefont {Teitelboim}(1972)}]{Teitelboim:1972ps}%
  \BibitemOpen
  \bibfield  {author} {\bibinfo {author} {\bibfnamefont {C.}~\bibnamefont {Teitelboim}},\ }\bibfield  {title} {\bibinfo {title} {{Nonmeasurability of the lepton number of a black hole}},\ }\href {https://doi.org/10.1007/BF02826050} {\bibfield  {journal} {\bibinfo  {journal} {Lett. Nuovo Cim.}\ }\textbf {\bibinfo {volume} {3S2}},\ \bibinfo {pages} {397} (\bibinfo {year} {1972})}\BibitemShut {NoStop}%
\bibitem [{\citenamefont {Bekenstein}(1972{\natexlab{a}})}]{Bekenstein:1971hc}%
  \BibitemOpen
  \bibfield  {author} {\bibinfo {author} {\bibfnamefont {J.~D.}\ \bibnamefont {Bekenstein}},\ }\bibfield  {title} {\bibinfo {title} {{Nonexistence of baryon number for static black holes}},\ }\href {https://doi.org/10.1103/PhysRevD.5.1239} {\bibfield  {journal} {\bibinfo  {journal} {Phys. Rev. D}\ }\textbf {\bibinfo {volume} {5}},\ \bibinfo {pages} {1239} (\bibinfo {year} {1972}{\natexlab{a}})}\BibitemShut {NoStop}%
\bibitem [{\citenamefont {Bekenstein}(1972{\natexlab{b}})}]{Bekenstein:1972ky}%
  \BibitemOpen
  \bibfield  {author} {\bibinfo {author} {\bibfnamefont {J.~D.}\ \bibnamefont {Bekenstein}},\ }\bibfield  {title} {\bibinfo {title} {{Nonexistence of baryon number for black holes. ii}},\ }\href {https://doi.org/10.1103/PhysRevD.5.2403} {\bibfield  {journal} {\bibinfo  {journal} {Phys. Rev. D}\ }\textbf {\bibinfo {volume} {5}},\ \bibinfo {pages} {2403} (\bibinfo {year} {1972}{\natexlab{b}})}\BibitemShut {NoStop}%
\bibitem [{\citenamefont {Berti}\ \emph {et~al.}(2015)\citenamefont {Berti} \emph {et~al.}}]{Berti:2015itd}%
  \BibitemOpen
  \bibfield  {author} {\bibinfo {author} {\bibfnamefont {E.}~\bibnamefont {Berti}} \emph {et~al.},\ }\bibfield  {title} {\bibinfo {title} {{Testing General Relativity with Present and Future Astrophysical Observations}},\ }\href {https://doi.org/10.1088/0264-9381/32/24/243001} {\bibfield  {journal} {\bibinfo  {journal} {Class. Quant. Grav.}\ }\textbf {\bibinfo {volume} {32}},\ \bibinfo {pages} {243001} (\bibinfo {year} {2015})},\ \Eprint {https://arxiv.org/abs/1501.07274} {arXiv:1501.07274 [gr-qc]} \BibitemShut {NoStop}%
\bibitem [{\citenamefont {Sotiriou}(2015)}]{Sotiriou:2015pka}%
  \BibitemOpen
  \bibfield  {author} {\bibinfo {author} {\bibfnamefont {T.~P.}\ \bibnamefont {Sotiriou}},\ }\bibfield  {title} {\bibinfo {title} {{Black Holes and Scalar Fields}},\ }\href {https://doi.org/10.1088/0264-9381/32/21/214002} {\bibfield  {journal} {\bibinfo  {journal} {Class. Quant. Grav.}\ }\textbf {\bibinfo {volume} {32}},\ \bibinfo {pages} {214002} (\bibinfo {year} {2015})},\ \Eprint {https://arxiv.org/abs/1505.00248} {arXiv:1505.00248 [gr-qc]} \BibitemShut {NoStop}%
\bibitem [{\citenamefont {Cardoso}\ and\ \citenamefont {Gualtieri}(2016)}]{Cardoso:2016ryw}%
  \BibitemOpen
  \bibfield  {author} {\bibinfo {author} {\bibfnamefont {V.}~\bibnamefont {Cardoso}}\ and\ \bibinfo {author} {\bibfnamefont {L.}~\bibnamefont {Gualtieri}},\ }\bibfield  {title} {\bibinfo {title} {{Testing the black hole {\textquoteleft}no-hair{\textquoteright} hypothesis}},\ }\href {https://doi.org/10.1088/0264-9381/33/17/174001} {\bibfield  {journal} {\bibinfo  {journal} {Class. Quant. Grav.}\ }\textbf {\bibinfo {volume} {33}},\ \bibinfo {pages} {174001} (\bibinfo {year} {2016})},\ \Eprint {https://arxiv.org/abs/1607.03133} {arXiv:1607.03133 [gr-qc]} \BibitemShut {NoStop}%
\bibitem [{\citenamefont {Herdeiro}(2023)}]{Herdeiro:2022yle}%
  \BibitemOpen
  \bibfield  {author} {\bibinfo {author} {\bibfnamefont {C.~A.~R.}\ \bibnamefont {Herdeiro}},\ }\bibfield  {title} {\bibinfo {title} {{Black Holes: On the Universality of the Kerr Hypothesis}},\ }\href {https://doi.org/10.1007/978-3-031-31520-6_8} {\bibfield  {journal} {\bibinfo  {journal} {Lect. Notes Phys.}\ }\textbf {\bibinfo {volume} {1017}},\ \bibinfo {pages} {315} (\bibinfo {year} {2023})},\ \Eprint {https://arxiv.org/abs/2204.05640} {arXiv:2204.05640 [gr-qc]} \BibitemShut {NoStop}%
\bibitem [{\citenamefont {Gair}\ \emph {et~al.}(2013)\citenamefont {Gair}, \citenamefont {Vallisneri}, \citenamefont {Larson},\ and\ \citenamefont {Baker}}]{Gair:2012nm}%
  \BibitemOpen
  \bibfield  {author} {\bibinfo {author} {\bibfnamefont {J.~R.}\ \bibnamefont {Gair}}, \bibinfo {author} {\bibfnamefont {M.}~\bibnamefont {Vallisneri}}, \bibinfo {author} {\bibfnamefont {S.~L.}\ \bibnamefont {Larson}},\ and\ \bibinfo {author} {\bibfnamefont {J.~G.}\ \bibnamefont {Baker}},\ }\bibfield  {title} {\bibinfo {title} {{Testing General Relativity with Low-Frequency, Space-Based Gravitational-Wave Detectors}},\ }\href {https://doi.org/10.12942/lrr-2013-7} {\bibfield  {journal} {\bibinfo  {journal} {Living Rev. Rel.}\ }\textbf {\bibinfo {volume} {16}},\ \bibinfo {pages} {7} (\bibinfo {year} {2013})},\ \Eprint {https://arxiv.org/abs/1212.5575} {arXiv:1212.5575 [gr-qc]} \BibitemShut {NoStop}%
\bibitem [{\citenamefont {Barack}\ \emph {et~al.}(2019)\citenamefont {Barack} \emph {et~al.}}]{Barack:2018yly}%
  \BibitemOpen
  \bibfield  {author} {\bibinfo {author} {\bibfnamefont {L.}~\bibnamefont {Barack}} \emph {et~al.},\ }\bibfield  {title} {\bibinfo {title} {{Black holes, gravitational waves and fundamental physics: a roadmap}},\ }\href {https://doi.org/10.1088/1361-6382/ab0587} {\bibfield  {journal} {\bibinfo  {journal} {Class. Quant. Grav.}\ }\textbf {\bibinfo {volume} {36}},\ \bibinfo {pages} {143001} (\bibinfo {year} {2019})},\ \Eprint {https://arxiv.org/abs/1806.05195} {arXiv:1806.05195 [gr-qc]} \BibitemShut {NoStop}%
\bibitem [{\citenamefont {Yunes}\ \emph {et~al.}(2024)\citenamefont {Yunes}, \citenamefont {Siemens},\ and\ \citenamefont {Yagi}}]{Yunes:2024lzm}%
  \BibitemOpen
  \bibfield  {author} {\bibinfo {author} {\bibfnamefont {N.}~\bibnamefont {Yunes}}, \bibinfo {author} {\bibfnamefont {X.}~\bibnamefont {Siemens}},\ and\ \bibinfo {author} {\bibfnamefont {K.}~\bibnamefont {Yagi}},\ }\bibfield  {title} {\bibinfo {title} {{Gravitational-Wave Tests of General Relativity with Ground-Based Detectors and Pulsar-Timing Arrays}},\ }\href@noop {} {\  (\bibinfo {year} {2024})},\ \Eprint {https://arxiv.org/abs/2408.05240} {arXiv:2408.05240 [gr-qc]} \BibitemShut {NoStop}%
\bibitem [{\citenamefont {Bekenstein}(1995)}]{Bekenstein:1995un}%
  \BibitemOpen
  \bibfield  {author} {\bibinfo {author} {\bibfnamefont {J.~D.}\ \bibnamefont {Bekenstein}},\ }\bibfield  {title} {\bibinfo {title} {{Novel \textquoteleft{}\textquoteleft{}no-scalar-hair\textquoteright{}\textquoteright{} theorem for black holes}},\ }\href {https://doi.org/10.1103/PhysRevD.51.R6608} {\bibfield  {journal} {\bibinfo  {journal} {Phys. Rev. D}\ }\textbf {\bibinfo {volume} {51}},\ \bibinfo {pages} {R6608} (\bibinfo {year} {1995})}\BibitemShut {NoStop}%
\bibitem [{\citenamefont {Sotiriou}\ and\ \citenamefont {Faraoni}(2012)}]{Sotiriou:2011dz}%
  \BibitemOpen
  \bibfield  {author} {\bibinfo {author} {\bibfnamefont {T.~P.}\ \bibnamefont {Sotiriou}}\ and\ \bibinfo {author} {\bibfnamefont {V.}~\bibnamefont {Faraoni}},\ }\bibfield  {title} {\bibinfo {title} {{Black holes in scalar-tensor gravity}},\ }\href {https://doi.org/10.1103/PhysRevLett.108.081103} {\bibfield  {journal} {\bibinfo  {journal} {Phys. Rev. Lett.}\ }\textbf {\bibinfo {volume} {108}},\ \bibinfo {pages} {081103} (\bibinfo {year} {2012})},\ \Eprint {https://arxiv.org/abs/1109.6324} {arXiv:1109.6324 [gr-qc]} \BibitemShut {NoStop}%
\bibitem [{\citenamefont {Hui}\ and\ \citenamefont {Nicolis}(2013)}]{Hui:2012qt}%
  \BibitemOpen
  \bibfield  {author} {\bibinfo {author} {\bibfnamefont {L.}~\bibnamefont {Hui}}\ and\ \bibinfo {author} {\bibfnamefont {A.}~\bibnamefont {Nicolis}},\ }\bibfield  {title} {\bibinfo {title} {{No-Hair Theorem for the Galileon}},\ }\href {https://doi.org/10.1103/PhysRevLett.110.241104} {\bibfield  {journal} {\bibinfo  {journal} {Phys. Rev. Lett.}\ }\textbf {\bibinfo {volume} {110}},\ \bibinfo {pages} {241104} (\bibinfo {year} {2013})},\ \Eprint {https://arxiv.org/abs/1202.1296} {arXiv:1202.1296 [hep-th]} \BibitemShut {NoStop}%
\bibitem [{\citenamefont {Herdeiro}\ and\ \citenamefont {Radu}(2015)}]{Herdeiro:2015waa}%
  \BibitemOpen
  \bibfield  {author} {\bibinfo {author} {\bibfnamefont {C.~A.~R.}\ \bibnamefont {Herdeiro}}\ and\ \bibinfo {author} {\bibfnamefont {E.}~\bibnamefont {Radu}},\ }\bibfield  {title} {\bibinfo {title} {{Asymptotically flat black holes with scalar hair: a review}},\ }\href {https://doi.org/10.1142/S0218271815420146} {\bibfield  {journal} {\bibinfo  {journal} {Int. J. Mod. Phys. D}\ }\textbf {\bibinfo {volume} {24}},\ \bibinfo {pages} {1542014} (\bibinfo {year} {2015})},\ \Eprint {https://arxiv.org/abs/1504.08209} {arXiv:1504.08209 [gr-qc]} \BibitemShut {NoStop}%
\bibitem [{\citenamefont {Kanti}\ \emph {et~al.}(1996)\citenamefont {Kanti}, \citenamefont {Mavromatos}, \citenamefont {Rizos}, \citenamefont {Tamvakis},\ and\ \citenamefont {Winstanley}}]{Kanti:1995vq}%
  \BibitemOpen
  \bibfield  {author} {\bibinfo {author} {\bibfnamefont {P.}~\bibnamefont {Kanti}}, \bibinfo {author} {\bibfnamefont {N.~E.}\ \bibnamefont {Mavromatos}}, \bibinfo {author} {\bibfnamefont {J.}~\bibnamefont {Rizos}}, \bibinfo {author} {\bibfnamefont {K.}~\bibnamefont {Tamvakis}},\ and\ \bibinfo {author} {\bibfnamefont {E.}~\bibnamefont {Winstanley}},\ }\bibfield  {title} {\bibinfo {title} {{Dilatonic black holes in higher curvature string gravity}},\ }\href {https://doi.org/10.1103/PhysRevD.54.5049} {\bibfield  {journal} {\bibinfo  {journal} {Phys. Rev. D}\ }\textbf {\bibinfo {volume} {54}},\ \bibinfo {pages} {5049} (\bibinfo {year} {1996})},\ \Eprint {https://arxiv.org/abs/hep-th/9511071} {arXiv:hep-th/9511071} \BibitemShut {NoStop}%
\bibitem [{\citenamefont {Pani}\ and\ \citenamefont {Cardoso}(2009)}]{Pani:2009wy}%
  \BibitemOpen
  \bibfield  {author} {\bibinfo {author} {\bibfnamefont {P.}~\bibnamefont {Pani}}\ and\ \bibinfo {author} {\bibfnamefont {V.}~\bibnamefont {Cardoso}},\ }\bibfield  {title} {\bibinfo {title} {{Are black holes in alternative theories serious astrophysical candidates? The Case for Einstein-Dilaton-Gauss-Bonnet black holes}},\ }\href {https://doi.org/10.1103/PhysRevD.79.084031} {\bibfield  {journal} {\bibinfo  {journal} {Phys. Rev. D}\ }\textbf {\bibinfo {volume} {79}},\ \bibinfo {pages} {084031} (\bibinfo {year} {2009})},\ \Eprint {https://arxiv.org/abs/0902.1569} {arXiv:0902.1569 [gr-qc]} \BibitemShut {NoStop}%
\bibitem [{\citenamefont {Yunes}\ and\ \citenamefont {Stein}(2011)}]{Yunes:2011we}%
  \BibitemOpen
  \bibfield  {author} {\bibinfo {author} {\bibfnamefont {N.}~\bibnamefont {Yunes}}\ and\ \bibinfo {author} {\bibfnamefont {L.~C.}\ \bibnamefont {Stein}},\ }\bibfield  {title} {\bibinfo {title} {{Non-Spinning Black Holes in Alternative Theories of Gravity}},\ }\href {https://doi.org/10.1103/PhysRevD.83.104002} {\bibfield  {journal} {\bibinfo  {journal} {Phys. Rev. D}\ }\textbf {\bibinfo {volume} {83}},\ \bibinfo {pages} {104002} (\bibinfo {year} {2011})},\ \Eprint {https://arxiv.org/abs/1101.2921} {arXiv:1101.2921 [gr-qc]} \BibitemShut {NoStop}%
\bibitem [{\citenamefont {Sotiriou}\ and\ \citenamefont {Zhou}(2014{\natexlab{a}})}]{Sotiriou:2013qea}%
  \BibitemOpen
  \bibfield  {author} {\bibinfo {author} {\bibfnamefont {T.~P.}\ \bibnamefont {Sotiriou}}\ and\ \bibinfo {author} {\bibfnamefont {S.-Y.}\ \bibnamefont {Zhou}},\ }\bibfield  {title} {\bibinfo {title} {{Black hole hair in generalized scalar-tensor gravity}},\ }\href {https://doi.org/10.1103/PhysRevLett.112.251102} {\bibfield  {journal} {\bibinfo  {journal} {Phys. Rev. Lett.}\ }\textbf {\bibinfo {volume} {112}},\ \bibinfo {pages} {251102} (\bibinfo {year} {2014}{\natexlab{a}})},\ \Eprint {https://arxiv.org/abs/1312.3622} {arXiv:1312.3622 [gr-qc]} \BibitemShut {NoStop}%
\bibitem [{\citenamefont {Sotiriou}\ and\ \citenamefont {Zhou}(2014{\natexlab{b}})}]{Sotiriou:2014pfa}%
  \BibitemOpen
  \bibfield  {author} {\bibinfo {author} {\bibfnamefont {T.~P.}\ \bibnamefont {Sotiriou}}\ and\ \bibinfo {author} {\bibfnamefont {S.-Y.}\ \bibnamefont {Zhou}},\ }\bibfield  {title} {\bibinfo {title} {{Black hole hair in generalized scalar-tensor gravity: An explicit example}},\ }\href {https://doi.org/10.1103/PhysRevD.90.124063} {\bibfield  {journal} {\bibinfo  {journal} {Phys. Rev. D}\ }\textbf {\bibinfo {volume} {90}},\ \bibinfo {pages} {124063} (\bibinfo {year} {2014}{\natexlab{b}})},\ \Eprint {https://arxiv.org/abs/1408.1698} {arXiv:1408.1698 [gr-qc]} \BibitemShut {NoStop}%
\bibitem [{\citenamefont {Maselli}\ \emph {et~al.}(2018)\citenamefont {Maselli}, \citenamefont {Pani}, \citenamefont {Cardoso}, \citenamefont {Abdelsalhin}, \citenamefont {Gualtieri},\ and\ \citenamefont {Ferrari}}]{Maselli:2017cmm}%
  \BibitemOpen
  \bibfield  {author} {\bibinfo {author} {\bibfnamefont {A.}~\bibnamefont {Maselli}}, \bibinfo {author} {\bibfnamefont {P.}~\bibnamefont {Pani}}, \bibinfo {author} {\bibfnamefont {V.}~\bibnamefont {Cardoso}}, \bibinfo {author} {\bibfnamefont {T.}~\bibnamefont {Abdelsalhin}}, \bibinfo {author} {\bibfnamefont {L.}~\bibnamefont {Gualtieri}},\ and\ \bibinfo {author} {\bibfnamefont {V.}~\bibnamefont {Ferrari}},\ }\bibfield  {title} {\bibinfo {title} {{Probing Planckian corrections at the horizon scale with LISA binaries}},\ }\href {https://doi.org/10.1103/PhysRevLett.120.081101} {\bibfield  {journal} {\bibinfo  {journal} {Phys. Rev. Lett.}\ }\textbf {\bibinfo {volume} {120}},\ \bibinfo {pages} {081101} (\bibinfo {year} {2018})},\ \Eprint {https://arxiv.org/abs/1703.10612} {arXiv:1703.10612 [gr-qc]} \BibitemShut {NoStop}%
\bibitem [{\citenamefont {Barausse}\ \emph {et~al.}(2020)\citenamefont {Barausse} \emph {et~al.}}]{Barausse:2020rsu}%
  \BibitemOpen
  \bibfield  {author} {\bibinfo {author} {\bibfnamefont {E.}~\bibnamefont {Barausse}} \emph {et~al.},\ }\bibfield  {title} {\bibinfo {title} {{Prospects for Fundamental Physics with LISA}},\ }\href {https://doi.org/10.1007/s10714-020-02691-1} {\bibfield  {journal} {\bibinfo  {journal} {Gen. Rel. Grav.}\ }\textbf {\bibinfo {volume} {52}},\ \bibinfo {pages} {81} (\bibinfo {year} {2020})},\ \Eprint {https://arxiv.org/abs/2001.09793} {arXiv:2001.09793 [gr-qc]} \BibitemShut {NoStop}%
\bibitem [{\citenamefont {Arun}\ \emph {et~al.}(2022)\citenamefont {Arun} \emph {et~al.}}]{LISA:2022kgy}%
  \BibitemOpen
  \bibfield  {author} {\bibinfo {author} {\bibfnamefont {K.~G.}\ \bibnamefont {Arun}} \emph {et~al.} (\bibinfo {collaboration} {LISA}),\ }\bibfield  {title} {\bibinfo {title} {{New horizons for fundamental physics with LISA}},\ }\href {https://doi.org/10.1007/s41114-022-00036-9} {\bibfield  {journal} {\bibinfo  {journal} {Living Rev. Rel.}\ }\textbf {\bibinfo {volume} {25}},\ \bibinfo {pages} {4} (\bibinfo {year} {2022})},\ \Eprint {https://arxiv.org/abs/2205.01597} {arXiv:2205.01597 [gr-qc]} \BibitemShut {NoStop}%
\bibitem [{\citenamefont {Colpi}\ \emph {et~al.}(2024)\citenamefont {Colpi} \emph {et~al.}}]{LISA:2024hlh}%
  \BibitemOpen
  \bibfield  {author} {\bibinfo {author} {\bibfnamefont {M.}~\bibnamefont {Colpi}} \emph {et~al.} (\bibinfo {collaboration} {LISA}),\ }\bibfield  {title} {\bibinfo {title} {{LISA Definition Study Report}},\ }\href@noop {} {\  (\bibinfo {year} {2024})},\ \Eprint {https://arxiv.org/abs/2402.07571} {arXiv:2402.07571 [astro-ph.CO]} \BibitemShut {NoStop}%
\bibitem [{\citenamefont {Abac}\ \emph {et~al.}(2025)\citenamefont {Abac} \emph {et~al.}}]{Abac:2025saz}%
  \BibitemOpen
  \bibfield  {author} {\bibinfo {author} {\bibfnamefont {A.}~\bibnamefont {Abac}} \emph {et~al.},\ }\bibfield  {title} {\bibinfo {title} {{The Science of the Einstein Telescope}},\ }\href@noop {} {\  (\bibinfo {year} {2025})},\ \Eprint {https://arxiv.org/abs/2503.12263} {arXiv:2503.12263 [gr-qc]} \BibitemShut {NoStop}%
\bibitem [{\citenamefont {Saravani}\ and\ \citenamefont {Sotiriou}(2019)}]{Saravani:2019xwx}%
  \BibitemOpen
  \bibfield  {author} {\bibinfo {author} {\bibfnamefont {M.}~\bibnamefont {Saravani}}\ and\ \bibinfo {author} {\bibfnamefont {T.~P.}\ \bibnamefont {Sotiriou}},\ }\bibfield  {title} {\bibinfo {title} {{Classification of shift-symmetric Horndeski theories and hairy black holes}},\ }\href {https://doi.org/10.1103/PhysRevD.99.124004} {\bibfield  {journal} {\bibinfo  {journal} {Phys. Rev. D}\ }\textbf {\bibinfo {volume} {99}},\ \bibinfo {pages} {124004} (\bibinfo {year} {2019})},\ \Eprint {https://arxiv.org/abs/1903.02055} {arXiv:1903.02055 [gr-qc]} \BibitemShut {NoStop}%
\bibitem [{\citenamefont {Thaalba}\ \emph {et~al.}(2023)\citenamefont {Thaalba}, \citenamefont {Antoniou},\ and\ \citenamefont {Sotiriou}}]{Thaalba:2022bnt}%
  \BibitemOpen
  \bibfield  {author} {\bibinfo {author} {\bibfnamefont {F.}~\bibnamefont {Thaalba}}, \bibinfo {author} {\bibfnamefont {G.}~\bibnamefont {Antoniou}},\ and\ \bibinfo {author} {\bibfnamefont {T.~P.}\ \bibnamefont {Sotiriou}},\ }\bibfield  {title} {\bibinfo {title} {{Black hole minimum size and scalar charge in shift-symmetric theories}},\ }\href {https://doi.org/10.1088/1361-6382/acdd42} {\bibfield  {journal} {\bibinfo  {journal} {Class. Quant. Grav.}\ }\textbf {\bibinfo {volume} {40}},\ \bibinfo {pages} {155002} (\bibinfo {year} {2023})},\ \Eprint {https://arxiv.org/abs/2211.05099} {arXiv:2211.05099 [gr-qc]} \BibitemShut {NoStop}%
\bibitem [{\citenamefont {Eichhorn}\ \emph {et~al.}(2025)\citenamefont {Eichhorn}, \citenamefont {Fernandes}, \citenamefont {Held},\ and\ \citenamefont {Silva}}]{Eichhorn:2023iab}%
  \BibitemOpen
  \bibfield  {author} {\bibinfo {author} {\bibfnamefont {A.}~\bibnamefont {Eichhorn}}, \bibinfo {author} {\bibfnamefont {P.~G.~S.}\ \bibnamefont {Fernandes}}, \bibinfo {author} {\bibfnamefont {A.}~\bibnamefont {Held}},\ and\ \bibinfo {author} {\bibfnamefont {H.~O.}\ \bibnamefont {Silva}},\ }\bibfield  {title} {\bibinfo {title} {{Breaking black-hole uniqueness at supermassive scales}},\ }\href {https://doi.org/10.1088/1361-6382/add3b6} {\bibfield  {journal} {\bibinfo  {journal} {Class. Quant. Grav.}\ }\textbf {\bibinfo {volume} {42}},\ \bibinfo {pages} {105006} (\bibinfo {year} {2025})},\ \Eprint {https://arxiv.org/abs/2312.11430} {arXiv:2312.11430 [gr-qc]} \BibitemShut {NoStop}%
\bibitem [{\citenamefont {Thaalba}\ \emph {et~al.}(2025)\citenamefont {Thaalba}, \citenamefont {Fernandes},\ and\ \citenamefont {Sotiriou}}]{Thaalba:2025ljh}%
  \BibitemOpen
  \bibfield  {author} {\bibinfo {author} {\bibfnamefont {F.}~\bibnamefont {Thaalba}}, \bibinfo {author} {\bibfnamefont {P.~G.~S.}\ \bibnamefont {Fernandes}},\ and\ \bibinfo {author} {\bibfnamefont {T.~P.}\ \bibnamefont {Sotiriou}},\ }\bibfield  {title} {\bibinfo {title} {{Supermassive black hole scalarization and effective field theory}},\ }\href@noop {} {\  (\bibinfo {year} {2025})},\ \Eprint {https://arxiv.org/abs/2506.21434} {arXiv:2506.21434 [gr-qc]} \BibitemShut {NoStop}%
\bibitem [{\citenamefont {Hui}\ \emph {et~al.}(2021)\citenamefont {Hui}, \citenamefont {Podo}, \citenamefont {Santoni},\ and\ \citenamefont {Trincherini}}]{Hui:2021cpm}%
  \BibitemOpen
  \bibfield  {author} {\bibinfo {author} {\bibfnamefont {L.}~\bibnamefont {Hui}}, \bibinfo {author} {\bibfnamefont {A.}~\bibnamefont {Podo}}, \bibinfo {author} {\bibfnamefont {L.}~\bibnamefont {Santoni}},\ and\ \bibinfo {author} {\bibfnamefont {E.}~\bibnamefont {Trincherini}},\ }\bibfield  {title} {\bibinfo {title} {{Effective Field Theory for the perturbations of a slowly rotating black hole}},\ }\href {https://doi.org/10.1007/JHEP12(2021)183} {\bibfield  {journal} {\bibinfo  {journal} {JHEP}\ }\textbf {\bibinfo {volume} {12}},\ \bibinfo {pages} {183}},\ \Eprint {https://arxiv.org/abs/2111.02072} {arXiv:2111.02072 [hep-th]} \BibitemShut {NoStop}%
\bibitem [{\citenamefont {Noller}\ \emph {et~al.}(2020)\citenamefont {Noller}, \citenamefont {Santoni}, \citenamefont {Trincherini},\ and\ \citenamefont {Trombetta}}]{Noller:2019chl}%
  \BibitemOpen
  \bibfield  {author} {\bibinfo {author} {\bibfnamefont {J.}~\bibnamefont {Noller}}, \bibinfo {author} {\bibfnamefont {L.}~\bibnamefont {Santoni}}, \bibinfo {author} {\bibfnamefont {E.}~\bibnamefont {Trincherini}},\ and\ \bibinfo {author} {\bibfnamefont {L.~G.}\ \bibnamefont {Trombetta}},\ }\bibfield  {title} {\bibinfo {title} {{Black Hole Ringdown as a Probe for Dark Energy}},\ }\href {https://doi.org/10.1103/PhysRevD.101.084049} {\bibfield  {journal} {\bibinfo  {journal} {Phys. Rev. D}\ }\textbf {\bibinfo {volume} {101}},\ \bibinfo {pages} {084049} (\bibinfo {year} {2020})},\ \Eprint {https://arxiv.org/abs/1911.11671} {arXiv:1911.11671 [gr-qc]} \BibitemShut {NoStop}%
\bibitem [{\citenamefont {Babichev}\ and\ \citenamefont {Deffayet}(2013)}]{Babichev:2013usa}%
  \BibitemOpen
  \bibfield  {author} {\bibinfo {author} {\bibfnamefont {E.}~\bibnamefont {Babichev}}\ and\ \bibinfo {author} {\bibfnamefont {C.}~\bibnamefont {Deffayet}},\ }\bibfield  {title} {\bibinfo {title} {{An introduction to the Vainshtein mechanism}},\ }\href {https://doi.org/10.1088/0264-9381/30/18/184001} {\bibfield  {journal} {\bibinfo  {journal} {Class. Quant. Grav.}\ }\textbf {\bibinfo {volume} {30}},\ \bibinfo {pages} {184001} (\bibinfo {year} {2013})},\ \Eprint {https://arxiv.org/abs/1304.7240} {arXiv:1304.7240 [gr-qc]} \BibitemShut {NoStop}%
\bibitem [{\citenamefont {Nicolis}\ \emph {et~al.}(2009)\citenamefont {Nicolis}, \citenamefont {Rattazzi},\ and\ \citenamefont {Trincherini}}]{Nicolis:2008in}%
  \BibitemOpen
  \bibfield  {author} {\bibinfo {author} {\bibfnamefont {A.}~\bibnamefont {Nicolis}}, \bibinfo {author} {\bibfnamefont {R.}~\bibnamefont {Rattazzi}},\ and\ \bibinfo {author} {\bibfnamefont {E.}~\bibnamefont {Trincherini}},\ }\bibfield  {title} {\bibinfo {title} {{The Galileon as a local modification of gravity}},\ }\href {https://doi.org/10.1103/PhysRevD.79.064036} {\bibfield  {journal} {\bibinfo  {journal} {Phys. Rev. D}\ }\textbf {\bibinfo {volume} {79}},\ \bibinfo {pages} {064036} (\bibinfo {year} {2009})},\ \Eprint {https://arxiv.org/abs/0811.2197} {arXiv:0811.2197 [hep-th]} \BibitemShut {NoStop}%
\bibitem [{\citenamefont {Luty}\ \emph {et~al.}(2003)\citenamefont {Luty}, \citenamefont {Porrati},\ and\ \citenamefont {Rattazzi}}]{Luty:2003vm}%
  \BibitemOpen
  \bibfield  {author} {\bibinfo {author} {\bibfnamefont {M.~A.}\ \bibnamefont {Luty}}, \bibinfo {author} {\bibfnamefont {M.}~\bibnamefont {Porrati}},\ and\ \bibinfo {author} {\bibfnamefont {R.}~\bibnamefont {Rattazzi}},\ }\bibfield  {title} {\bibinfo {title} {{Strong interactions and stability in the DGP model}},\ }\href {https://doi.org/10.1088/1126-6708/2003/09/029} {\bibfield  {journal} {\bibinfo  {journal} {JHEP}\ }\textbf {\bibinfo {volume} {09}},\ \bibinfo {pages} {029}},\ \Eprint {https://arxiv.org/abs/hep-th/0303116} {arXiv:hep-th/0303116} \BibitemShut {NoStop}%
\bibitem [{\citenamefont {Serra}\ \emph {et~al.}(2022)\citenamefont {Serra}, \citenamefont {Serra}, \citenamefont {Trincherini},\ and\ \citenamefont {Trombetta}}]{Serra:2022pzl}%
  \BibitemOpen
  \bibfield  {author} {\bibinfo {author} {\bibfnamefont {F.}~\bibnamefont {Serra}}, \bibinfo {author} {\bibfnamefont {J.}~\bibnamefont {Serra}}, \bibinfo {author} {\bibfnamefont {E.}~\bibnamefont {Trincherini}},\ and\ \bibinfo {author} {\bibfnamefont {L.~G.}\ \bibnamefont {Trombetta}},\ }\bibfield  {title} {\bibinfo {title} {{Causality constraints on black holes beyond GR}},\ }\href {https://doi.org/10.1007/JHEP08(2022)157} {\bibfield  {journal} {\bibinfo  {journal} {JHEP}\ }\textbf {\bibinfo {volume} {08}},\ \bibinfo {pages} {157}},\ \Eprint {https://arxiv.org/abs/2205.08551} {arXiv:2205.08551 [hep-th]} \BibitemShut {NoStop}%
\bibitem [{\citenamefont {Kleihaus}\ \emph {et~al.}(2016)\citenamefont {Kleihaus}, \citenamefont {Kunz}, \citenamefont {Mojica},\ and\ \citenamefont {Radu}}]{Kleihaus:2015aje}%
  \BibitemOpen
  \bibfield  {author} {\bibinfo {author} {\bibfnamefont {B.}~\bibnamefont {Kleihaus}}, \bibinfo {author} {\bibfnamefont {J.}~\bibnamefont {Kunz}}, \bibinfo {author} {\bibfnamefont {S.}~\bibnamefont {Mojica}},\ and\ \bibinfo {author} {\bibfnamefont {E.}~\bibnamefont {Radu}},\ }\bibfield  {title} {\bibinfo {title} {{Spinning black holes in Einstein{\textendash}Gauss-Bonnet{\textendash}dilaton theory: Nonperturbative solutions}},\ }\href {https://doi.org/10.1103/PhysRevD.93.044047} {\bibfield  {journal} {\bibinfo  {journal} {Phys. Rev. D}\ }\textbf {\bibinfo {volume} {93}},\ \bibinfo {pages} {044047} (\bibinfo {year} {2016})},\ \Eprint {https://arxiv.org/abs/1511.05513} {arXiv:1511.05513 [gr-qc]} \BibitemShut {NoStop}%
\bibitem [{\citenamefont {Delgado}\ \emph {et~al.}(2020)\citenamefont {Delgado}, \citenamefont {Herdeiro},\ and\ \citenamefont {Radu}}]{Delgado:2020rev}%
  \BibitemOpen
  \bibfield  {author} {\bibinfo {author} {\bibfnamefont {J.~F.~M.}\ \bibnamefont {Delgado}}, \bibinfo {author} {\bibfnamefont {C.~A.~R.}\ \bibnamefont {Herdeiro}},\ and\ \bibinfo {author} {\bibfnamefont {E.}~\bibnamefont {Radu}},\ }\bibfield  {title} {\bibinfo {title} {{Spinning black holes in shift-symmetric Horndeski theory}},\ }\href {https://doi.org/10.1007/JHEP04(2020)180} {\bibfield  {journal} {\bibinfo  {journal} {JHEP}\ }\textbf {\bibinfo {volume} {04}},\ \bibinfo {pages} {180}},\ \Eprint {https://arxiv.org/abs/2002.05012} {arXiv:2002.05012 [gr-qc]} \BibitemShut {NoStop}%
\bibitem [{\citenamefont {Creminelli}\ \emph {et~al.}(2020)\citenamefont {Creminelli}, \citenamefont {Loayza}, \citenamefont {Serra}, \citenamefont {Trincherini},\ and\ \citenamefont {Trombetta}}]{Creminelli:2020lxn}%
  \BibitemOpen
  \bibfield  {author} {\bibinfo {author} {\bibfnamefont {P.}~\bibnamefont {Creminelli}}, \bibinfo {author} {\bibfnamefont {N.}~\bibnamefont {Loayza}}, \bibinfo {author} {\bibfnamefont {F.}~\bibnamefont {Serra}}, \bibinfo {author} {\bibfnamefont {E.}~\bibnamefont {Trincherini}},\ and\ \bibinfo {author} {\bibfnamefont {L.~G.}\ \bibnamefont {Trombetta}},\ }\bibfield  {title} {\bibinfo {title} {{Hairy Black-holes in Shift-symmetric Theories}},\ }\href {https://doi.org/10.1007/JHEP08(2020)045} {\bibfield  {journal} {\bibinfo  {journal} {JHEP}\ }\textbf {\bibinfo {volume} {08}},\ \bibinfo {pages} {045}},\ \Eprint {https://arxiv.org/abs/2004.02893} {arXiv:2004.02893 [hep-th]} \BibitemShut {NoStop}%
\bibitem [{\citenamefont {Yale}\ and\ \citenamefont {Padmanabhan}(2011)}]{Yale:2011usf}%
  \BibitemOpen
  \bibfield  {author} {\bibinfo {author} {\bibfnamefont {A.}~\bibnamefont {Yale}}\ and\ \bibinfo {author} {\bibfnamefont {T.}~\bibnamefont {Padmanabhan}},\ }\bibfield  {title} {\bibinfo {title} {{Structure of Lanczos-Lovelock Lagrangians in Critical Dimensions}},\ }\href {https://doi.org/10.1007/s10714-011-1146-1} {\bibfield  {journal} {\bibinfo  {journal} {Gen. Rel. Grav.}\ }\textbf {\bibinfo {volume} {43}},\ \bibinfo {pages} {1549} (\bibinfo {year} {2011})},\ \Eprint {https://arxiv.org/abs/1008.5154} {arXiv:1008.5154 [gr-qc]} \BibitemShut {NoStop}%
\bibitem [{\citenamefont {Dar}\ \emph {et~al.}(2019)\citenamefont {Dar}, \citenamefont {De~Rham}, \citenamefont {Deskins}, \citenamefont {Giblin},\ and\ \citenamefont {Tolley}}]{Dar:2018dra}%
  \BibitemOpen
  \bibfield  {author} {\bibinfo {author} {\bibfnamefont {F.}~\bibnamefont {Dar}}, \bibinfo {author} {\bibfnamefont {C.}~\bibnamefont {De~Rham}}, \bibinfo {author} {\bibfnamefont {J.~T.}\ \bibnamefont {Deskins}}, \bibinfo {author} {\bibfnamefont {J.~T.}\ \bibnamefont {Giblin}},\ and\ \bibinfo {author} {\bibfnamefont {A.~J.}\ \bibnamefont {Tolley}},\ }\bibfield  {title} {\bibinfo {title} {{Scalar Gravitational Radiation from Binaries: Vainshtein Mechanism in Time-dependent Systems}},\ }\href {https://doi.org/10.1088/1361-6382/aaf5e8} {\bibfield  {journal} {\bibinfo  {journal} {Class. Quant. Grav.}\ }\textbf {\bibinfo {volume} {36}},\ \bibinfo {pages} {025008} (\bibinfo {year} {2019})},\ \Eprint {https://arxiv.org/abs/1808.02165} {arXiv:1808.02165 [hep-th]} \BibitemShut {NoStop}%
\bibitem [{\citenamefont {S\"anger}\ \emph {et~al.}(2024)\citenamefont {S\"anger} \emph {et~al.}}]{Sanger:2024axs}%
  \BibitemOpen
  \bibfield  {author} {\bibinfo {author} {\bibfnamefont {E.~M.}\ \bibnamefont {S\"anger}} \emph {et~al.},\ }\bibfield  {title} {\bibinfo {title} {{Tests of General Relativity with GW230529: a neutron star merging with a lower mass-gap compact object}},\ }\href@noop {} {\  (\bibinfo {year} {2024})},\ \Eprint {https://arxiv.org/abs/2406.03568} {arXiv:2406.03568 [gr-qc]} \BibitemShut {NoStop}%
\end{thebibliography}%
\end{document}